\documentclass[12pt]{article}
\usepackage[latin9]{inputenc}
\usepackage{amsmath,amsfonts,amssymb,amsthm,amstext,amscd,array}
\usepackage[table, dvipsnames]{xcolor}

\usepackage{soul}
\usepackage{tikz}
\usepackage{mathrsfs}
\usepackage{bbm}
\usepackage{bm}
\usepackage{indentfirst}
\usepackage[arrow,matrix,curve]{xy}
\usepackage{hyperref}

\renewcommand{\thefootnote}{\fnsymbol{footnote}}
\newcommand{\newsection}{\setcounter{equation}{0}\section}

\makeatother

\begin{document}
\begin{titlepage}
\vspace{3cm}
\baselineskip=24pt

\begin{center}
\textbf{\LARGE{}Asymptotic symmetries of three-dimensional Chern-Simons gravity for the Maxwell algebra}
\par\end{center}{\LARGE \par}

\begin{center}
	\vspace{1cm}
	\textbf{Patrick Concha}$^{\ast}$,
	\textbf{Nelson Merino}$^{\dag}$,
	\textbf{Olivera Miskovic}$^{\ast}$,
	\textbf{Evelyn Rodríguez}$^{\ddag}$,
	\\[3mm]\textbf{Patricio Salgado-Rebolledo}$^{\sharp}$
	\textbf{and} \textbf{Omar Valdivia}$^{\flat }$
	\footnotesize
	\\[5mm]
	$^{\flat }$\textit{Facultad de Ingeniería y Arquitectura,}\\
	\textit{ Universidad Arturo Prat, Iquique-Chile.}
	\\
	$^{\ast}$\textit{Instituto
		de Física, Pontificia Universidad Católica de Valparaíso, }\\
	\textit{ Casilla 4059, Valparaiso-Chile.}
	\\
	$^{\dag}$\textit{APC, CNRS-Universitè Paris 7, 75205 Paris CEDEX 13,
		France.} \\\textit{ }
	$^{\ddag}$\textit{Departamento de Ciencias, Facultad de Artes Liberales,} \\
	\textit{Universidad Adolfo Ibáñez, Viña del Mar-Chile.} \\
	$^{\sharp}$\textit{Facultad
		de Ingeniería y Ciencias \& UAI Physics Center, Universidad Adolfo Ibáñez,}\\
	\textit{Avda. Diagonal Las Torres 2460, Santiago-Chile.} \\[5mm]
	\footnotesize
	\texttt{patrick.concha@pucv.cl},
	\texttt{nemerino@gmail.com},
	\texttt{olivera.miskovic@pucv.cl},\\
	\texttt{evelyn.rodriguez@edu.uai.cl},
	\texttt{patricio.salgado@edu.uai.cl},
	\texttt{ovaldivi@unap.cl}
	\par\end{center}
\vskip 15pt
\begin{abstract}
\noindent
We study a three-dimensional Chern-Simons gravity theory based on the Maxwell algebra. We find that the boundary dynamics is described by an enlargement and deformation of the  $\mathfrak{bms}_3$ algebra with three independent central charges. This symmetry arises from a gravity action invariant under the local Maxwell group and is characterized by presence of Abelian generators which modify the commutation relations of the super-translations in the standard $\mathfrak{bms}_3$ algebra. Our analysis is based on the charge algebra of the theory in the BMS gauge, which includes the known solutions of standard asymptotically flat case. The field content of the theory is different than the one of General Relativity, but it includes all its geometries as particular solutions. In this line, we also study the stationary solutions of the theory in ADM form and we show that the vacuum energy and the vacuum angular momentum of the stationary configuration are influenced by the presence of the gravitational Maxwell field.

\end{abstract}
\end{titlepage} \setcounter{page}{2}

\newpage{}

{\baselineskip=12pt \tableofcontents{}}

\global\long\def\thefootnote{\arabic{footnote}}
 \setcounter{footnote}{0}

\newsection{Introduction}

In the last decades, three-dimensional gravity has become an interesting testing ground for understanding diverse intricate aspects of the gravitational interaction and underlying laws of quantum gravity. Even in the simplest case of General Relativity (GR) when the theory exhibits no propagating local degrees of freedom, it shares many properties with  higher-dimensional models, such as the existence of black hole solutions and its thermodynamics \cite{Banados:1992wn}.

In the context of the AdS/CFT correspondence, the study of three-dimensional gravity with negative cosmological constant has led to a number of remarkable results due to its rich boundary dynamics, starting from a striking discovery that the asymptotic symmetry of three-dimensional GR is given by a central extension of the conformal algebra in two dimensions \cite{Brown:1986nw}.
As the holographic principle is believed to be true beyond the AdS/CFT scenario, extensions of these results to other asymptotics were considered as worth to be studied. For example, extensions of the original results have been found in diverse contexts in Refs.~\cite{Henneaux:2009pw,Skenderis:2009nt,Afshar:2011qw,Sinha:2010ai,Compere:2013bya,Troessaert:2013fma,Perez:2016vqo,Grumiller:2016pqb}.

In asymptotically flat spacetimes, the asymptotic symmetry of GR~\cite{Barnich:2006av} is described by the three-dimensional version of the
Bondi-Metzner-Sachs (BMS) group \cite{Bondi:1962px,Sachs:1962zza,Ashtekar:1996cd,Barnich:2010eb}. The associated $\mathfrak{bms}_3$ algebra can be obtained from the conformal algebra by performing a flat limit through an Inönü-Wigner contraction, in a similar way as the Poincaré algebra follows from the AdS one \cite{Bagchi:2012cy,Barnich:2012aw,Gonzalez:2013oaa, Barnich:2013yka, Costa:2013vza,Fareghbal:2013ifa,Krishnan:2013wta}.  It has also been found as a conformal extension of Carroll group in the context on non-relativistic symmetries~\cite{Duval:2014uva}. Cosmological solutions of flat Einstein gravity can be recovered from the BTZ black hole through the flat limit, whose thermodynamics is well-defined and its entropy is given by a Cardy-like formula~\cite{Barnich:2012xq}.
Recent studies and applications of the $\mathfrak{bms}_{3}$
(super)algebra in the context of (super)gravity can be found, for example, in Refs.~\cite{Barnich:2014cwa,Basu:2017aqn,Fuentealba:2017fck}\footnote{Interestingly, the $\mathfrak{bms}_{3}$ algebra is not particular to gravitational systems. In fact, it can be realized as a symmetry of the three-dimensional Klein-Gordon action on flat space \cite{Batlle:2017llu}. }. More general asymptotically flat boundary conditions have been introduced in Ref.~\cite{Detournay:2016sfv}, giving rise to the
asymptotic symmetry described by a semi-direct product of a $\mathfrak{bms}_{3}$
algebra and two $\mathfrak{u}(1)$ current algebras. Such asymptotic symmetry has also been studied in Ref.~\cite{Setare:2017mry}
and can be recovered as a flat limit of the enhanced asymptotic symmetry of AdS$_{3}$ spacetime introduced in Ref.~\cite{Troessaert:2013fma}.
This work is devoted to exploring the presence of additional fields and their effect to the both bulk and boundary gravitational dynamics.

Let us emphasize that an advantage of studying gravity based on extended symmetries relies on the fact that any matter field interaction can be introduced via the minimal coupling, following only the gauge principle. This is in contrast to standard gravitational models where non-minimal couplings, exotic theories and higher dimensions are needed to obtain richer effective theories. An extended symmetry enlarges dynamical properties of a theory in a much simpler way, at the same time opening a possibility to obtain new solutions with interesting physical content.

Extended asymptotic symmetries  usually involve modifications of the bulk dynamics and arise from theories with a different field content with respect to GR. For example, it is known that the presence of a constant classical electromagnetic field background in Minkowski space changes Poincar\'e symmetries and gives Maxwell algebra which contains additional generators and modifies the Poincar\'e algebra \cite{Schrader:1972zd, Bacry1970, Gomis:2017cmt}.
In any dimension, the Maxwell algebra can be obtained as an Inönü-Wigner contraction of the AdS-Lorentz algebra. On the other hand, it can be derived as an S-expansion \cite{Izaurieta:2006zz} of the AdS algebra \cite{Salgado:2014qqa}.

Recently, the authors of \cite{deAzcarraga:2010sw} have shown that the Maxwell algebra allows to introduce an alternative effective cosmological constant term to gravity in four dimensions, that includes a contribution from the additional vector fields. On the other hand, in higher dimensions, the use of the Maxwell gauge symmetry enables to obtain new infinite high spin Maxwell multiplets with their massless
components coupled to each other through constant EM background \cite{Fedoruk:2013sna}. These applications of the Maxwell algebra show that deformed and enlarged symmetries can come as physical consequences of having additional gauge fields in a gravity theory, or from the fact that extended gauge symmetries naturally occur if gravity is an effective theory coming from extra dimensions. Other particular applications in the framework of the dynamics are yet to be explored. In this work we will focus to its asymptotic counterpart.

In three dimensions, a gravity theory invariant under the Maxwell algebra can be naturally constructed as a Chern-Simons (CS) action. This theory turns out to be given by Einstein-Hilbert (EH) term without the cosmological constant plus a non back-reacting matter field coupled to the geometry~\cite{Salgado:2014jka}. In the last years, the Maxwell (super)symmetry and its generalizations have been developed in the context of (super)gravity theories leading to a number of interesting results ~\cite{Bonanos:2009wy,Hoseinzadeh:2014bla,Concha:2015woa,Concha:2016hbt,Concha:2016zdb,Caroca:2017izc, Aviles:2018jzw}.

In this paper, we present a novel asymptotic symmetry of a gravitational theory without the cosmological constant by analyzing the asymptotic structure of the three-dimensional CS gravity invariant under the action of the Maxwell group \cite{Salgado:2014jka}.
A motivation to do so is twofold. On one hand, there has been renewed interest in asymptotic symmetries at null infinity because they can be related to the nontrivial symmetries of the quantum S-matrix~\cite{Strominger:2013lka,Strominger:2013jfa}, and their Ward identities to the soft theorems. On the other hand, a presence of an additional field coupled to gravity does not change only the asymptotic sector, but also the vacuum of the theory. In the framework of Chern-Simons (CS) gravity with extended gauge symmetry, it is possible to have a Minkowski vacuum that includes non-vanishing extra fields, which changes its total energy and the angular momentum. Furthermore, a richer vacuum structure opens the possibility of having a wider space of topological solutions in this theory.

 The asymptotic symmetry in this case is described by a deformed $\mathfrak{bms}_{3}$ algebra, which has been recently introduced in \cite{Caroca:2017onr} as an S-expansion of the Virasoro algebra. This symmetry contains an infinite-dimensional Abelian ideal spanned by generators $\mathcal{Z}_{m}$,
which modifies the $\mathfrak{bms}_{3}$ commutation relations of the supertranslations $\mathcal{P}_{m}$, such that
\begin{equation}\label{compmpn}
\left[\mathcal{P}_{m},\mathcal{P}_{n}\right]=\left(m-n\right)\mathcal{Z}_{m+n}+\frac{c_{3}}{12}\left(m^{3}-m\right)\delta_{m+n,0}\,.
\end{equation}
The presence of an additional set of generators implies the existence
of an extra central charge $c_{3}$. In the present work,  we show that the new central
charge is related to the coupling constant $\alpha_{2}$ arising from the Maxwell
sector of the gravitational action. The Maxwell algebra is a finite subalgebra of \eqref{compmpn} and the fact that it can be obtained as a deformation and enlargement of the Poincaré algebra is inherited from the asymptotic symmetries.

The paper is organized as follows: In section~\ref{two},  we
make a brief summary of the three-dimensional CS gravity invariant under
the Maxwell symmetry. Considering asymptotically flat geometries with time-like boundary, we present stationary solutions using the ADM ansatz, and when the boundary is the null like, we discuss the extended solutions of the theory using the BMS gauge. In both cases we provide boundary conditions which keep the action stationary on-shell. In section \ref{three}, we focus on the solutions in the BMS gauge and show that a deformed $\mathfrak{bms}_{3}$ algebra corresponds to the asymptotic symmetry
of the theory. Section \ref{four} is devoted to discussion and possible follow-ups.

\newsection{Three-dimensional gravity with local Maxwell symmetry}\label{two}

We are interested in a gravity theory invariant under an enlargement of
Poincaré group, such that General Relativity (GR) is recovered as a particular case. A minimal non-trivial set of symmetry transformations fulfilling these conditions is described by the Maxwell
group \cite{Schrader:1972zd,Bacry1970,Gomis:2017cmt}. In three-dimensions, the Maxwell algebra is generated by translations $P_{a}$, Lorentz transformations $J_{a}$ and an Abelian ideal of generators $Z_{a}$ satisfying the commutation relations\begin{equation}
\begin{tabular}{ll}
 \ensuremath{\left[J_{a},J_{b}\right]=\epsilon_{abc}J^{c}\,,}  \qquad \medskip &  \ensuremath{\left[P_{a},P_{b}\right]=\epsilon_{abc}Z^{c}\,},\\
 \ensuremath{\left[J_{a},Z_{b}\right]=\epsilon_{abc}Z^{c}\,,}   \medskip &  \ensuremath{\left[Z_{a},Z_{b}\right]=0}\,,\\
 \ensuremath{\left[J_{a},P_{b}\right]=\epsilon_{abc}P^{c}}\,,    &  \ensuremath{\left[Z_{a},P_{b}\right]=0}\,,
\end{tabular}\ \label{cs2}
\end{equation}
where $a,b,\ldots=0,1,2$ are Lorentz indices raised and lowered with
the Minkowski metric $\eta_{ab}$, and $\epsilon_{abc}$ is the three-dimensional Levi-Civita tensor. All generators are anti-Hermitean. Note that the generators $(P_a, J_a)$ do not form a Poincaré subalgebra of (\ref{cs2}) because the translations
$P_a$ do not commute. At first sight, it looks like $Z_a$ should be interpreted as the Poincaré translations because the set $(Z_a, J_a)$ is closed, forming a Poincaré subalgebra. However, as we shall see below, identifying $P_a$ with the translational generator gives a good gravitational dynamics, where GR is reproduced in a particular limit.

It is well-known that the Maxwell algebra can be obtained by an S-expansion of the
anti-de Sitter algebra in any dimension~\cite{Salgado:2014qqa}.
One of the main virtues of the S-expansion procedure is that it provides the full set of invariant tensors of the expanded algebra~\cite{Izaurieta:2006zz}. Having the invariant tensor is the key ingredient for studying a CS gravitational theory. For the present algebra (\ref{cs2}), the
relevant tensor in three-dimensional spacetime has rank 2, and its components are given by
\begin{equation}
\begin{tabular}{ll}
 \ensuremath{\left\langle J_{a}J_{b}\right\rangle =\alpha_{0}\,\eta_{ab}\,,}\qquad  \medskip&  \ensuremath{\left\langle P_{a}P_{b}\right\rangle =\alpha_{2}\,\eta_{ab}\,,}\\
 \ensuremath{\left\langle J_{a}P_{b}\right\rangle =\alpha_{1}\,\eta_{ab}\,,} \medskip &  \ensuremath{\left\langle Z_{a}Z_{b}\right\rangle =0}\,,\\
 \ensuremath{\left\langle J_{a}Z_{b}\right\rangle =\alpha_{2}\,\eta_{ab}\,,}  &  \ensuremath{\left\langle Z_{a}P_{b}\right\rangle =0}\,,
\end{tabular}\ \label{cs3}
\end{equation}
where $\left\langle \cdots \right\rangle$ stands for a non-degenerate invariant symmetric bilinear form and $\alpha_{0},\alpha_{1}$ and $\alpha_{2}$ are real
dimensionless constants. The fundamental field associated to the Maxwell algebra is the one-form
gauge potential
\begin{equation}
A=e^{a}P_{a}+\omega^{a}J_{a}+\sigma^{a}Z_{a}\,,\label{cs1}
\end{equation}
whose components are the vielbein $e^{a}(x)$, the spin connection $\omega^{a}(x)$
and  the gravitational Maxwell gauge field $\sigma^{a}(x)$. The dynamics of the field $A$ in three dimensions is described by the CS action,
\begin{equation}
I[A]=\frac{k}{4\pi}\int\limits _{\mathcal{M}}\left\langle AdA+\frac{2}{3}\,A^{3}\right\rangle\,,
\end{equation}
where a three-dimensional manifold ${\cal M}$ has non-trivial boundary. The Maxwell symmetry
is guaranteed, up to a total derivative, by the use of the Maxwell gauge field (\ref{cs1}),
the invariant tensor (\ref{cs3}) and the algebra (\ref{cs2}). The level
of the theory, $k=\frac{1}{4G}$, is related to the gravitational
constant $G$. For the sake of simplicity, we omit writing the wedge product between the forms. As shown in Ref.~\cite{Salgado:2014jka}, the CS action with Maxwell symmetry reads
\begin{align}
I[A] =& \frac{k}{4\pi}\int\limits_{\mathcal{M}}\left[ \alpha_{0}\left(\omega^{a}d\omega_{a} +\frac{1}{3}\,\epsilon^{abc}\omega_{a}\omega_{b}\omega_{c}\right) +2\alpha_{1}R_{a}e^{a}\right. \nonumber \\
 & +\left. \alpha_{2}\left(T^{a}e_{a}+2R^{a}\sigma_{a}\right)-d\left(\alpha_{1}\omega^{a}e_{a} +\alpha_{2}\omega^{a}\sigma_{a}\right)\rule{0pt}{15pt}\right]\,.\label{cs4}
\end{align}
The first term in the action is the gravitational CS term with the coupling
constant $\alpha _0$. Next term is the EH one, so that its
coupling can be normalized to $\alpha _1=1$. The last term, with the coupling constant $\alpha _2$, gives the dynamics to the gravitational Maxwell field
and also contributes to the dynamics of the other fields.

The action (\ref{cs4}) is invariant, up to boundary terms, under
the action of the infinitesimal gauge transformations $\delta A=d\Lambda+[A,\Lambda]$. In terms of components, the local gauge parameter reads  $\Lambda=\varepsilon^{a}(x)P_{a}+\chi^{a}(x)J_{a}+\gamma^{a}(x)Z_{a}$, and
the gauge fields change as
\begin{align}
\delta_{\Lambda}e^{a} & =D\varepsilon^{a}-\epsilon^{abc}\chi_{b}e_{c}\,,\nonumber \\
\delta_{\Lambda}\omega^{a} & =D\chi^{a}\,,\label{cs7}\\
\delta_{\Lambda}\sigma^{a} & =D\gamma^{a}+\epsilon^{abc}\left(e_{b}\varepsilon_{c}-\chi_{b}\sigma_{c}\right)\,,\nonumber
\end{align}
where the Lorentz covariant derivative is defined by $Dv^{a}=dv^{a}+\epsilon^{abc}\omega_{b}v_{c}$.

CS theories are also invariant under general coordinate transformations $\delta x^{\mu }=\xi ^{\mu }(x)$ by construction. They act on the connection one-form as a
Lie derivative, $\mathcal{L}_{\xi }A=\partial _{\mu }\xi ^{\nu }A_{\nu }+\xi^{\nu }\partial _{\nu }A_{\mu }$, and are on-shell equivalent to gauge transformations with the local parameter $\Lambda =\iota _{\xi }A=\xi^\mu A_\mu$. Here, $\iota_{\xi}$ denotes the contraction with respect to the vector field $\xi$. In particular, in case of considered gravitational theory, a field-dependent gauge transformation with the parameters $\varepsilon ^{a}=\iota_{\xi }e^{a}$, $\chi ^{a}=\iota_{\xi }\omega ^{a}$ and $\gamma ^{a}=\iota_{\xi }\sigma ^{a}$
becomes a general coordinate transformation,
\begin{eqnarray}
\mathcal{L}_{\xi }e^{a} &=&\delta _{\Lambda }e^{a}+\iota _{\xi }T^{a}\,, \notag \\
\mathcal{L}_{\xi }\omega ^{a} &=&\delta _{\Lambda }\omega ^{a}+\iota_{\xi}R^{a}\,,  \label{Lie} \\
\mathcal{L}_{\xi }\sigma ^{a} &=&\delta _{\Lambda }\sigma ^{a}+\iota_{\xi}\left( D\sigma ^{a}+\frac{1}{2}\,\epsilon ^{abc}e_{b}e_{c}\right) \,,
\notag
\end{eqnarray}
provided the field equations are satisfied. Extremization of the action (\ref{cs4}) gives rise to the following
equations of motion,
\begin{eqnarray}
\delta e^{a} & : & \qquad0=\alpha_{1}R_{a}+\alpha_{2}T_{a}\,,\nonumber \\
\delta\omega^{a} & : & \qquad0=\alpha_{0}R_{a}+\alpha_{1}T_{a}+\alpha_{2}\left(D\sigma_{a}+\frac{1}{2}\,\epsilon_{abc}e^{b}e^{c}\right)\,,\label{cs5-1}\\
\delta\sigma^{a} & : & \qquad0=\alpha_{2}R_{a}\,,\nonumber
\end{eqnarray}
where the curvature and torsion two-forms are $R^{a} =d\omega^{a}+\frac{1}{2}\,\epsilon^{abc}\omega_{b}\omega_{c}$ and $T^a =De^a$, respectively.
In the limit $\alpha_{2}=0$, the equations of motion of GR are recovered without
($\alpha_{0}=0$) and with ($\alpha_{0}\neq0$) the gravitational
CS term. When $\alpha_{2}\neq0$, the above
equations can be equivalently written as
\begin{align}
T^{a} & =0\,,\nonumber \\
R^{a} & =0\,,\label{cs5'-1}\\
D\sigma^{a}+\frac{1}{2}\,\epsilon^{abc}e_{b}e_{c} & =0\,,\nonumber
\end{align}
showing that eqs.~(\ref{Lie}) indeed give $\mathcal{L}_{\xi }= \delta _{\Lambda }$ on-shell. As a consequence, conserved charges can be calculated using any of these two transformation laws \cite{Banados:1994tn}.

Similarly to GR, the geometries described by the
equations of motion (\ref{cs5'-1}) are Riemannian (torsionless) and
locally flat. A difference with respect to GR is that
the gravitational Maxwell field, $\sigma^{a}$, does not vanish on-shell when $\alpha_2\neq0$ because $D\sigma_a$ must be proportional to a constant. In particular, contracting the last equation in (\ref{cs5'-1}) by $e_a$, we get on-shell that $e_{a}D\sigma^{a}$ is the 3D volume element, and therefore a strictly non-vanishing quantity. Furthermore, $\sigma^{a}$ couples
to the geometry through the interaction with $\omega^a$ and $e^b$, that leads to new effects compared to GR.
For example, it modifies the asymptotic sector of the spacetime and the asymptotic charges of the solutions.

In order to have a well-defined action, one has to ensure that the
boundary terms in $\delta I$ vanish on-shell upon imposing suitable boundary
conditions on the fields. Indeed, the on-shell variation of
the CS action is a surface term,
\begin{equation}
\delta I_{\mathrm{on-shell}}=\frac{k}{4\pi}\int\limits _{\partial\mathcal{M}}\left\langle \delta A\,A\right\rangle \,,\label{Var}
\end{equation}
which, in our case, takes the form
\begin{equation}
\delta I_{\mathrm{on-shell}}=\frac{k}{4\pi}\int\limits _{\partial\mathcal{M}}\left[\delta e^{a}\left(\omega_{a}+\alpha_{2}e_{a}\right)+\delta\omega^{a}\left(e_{a}+\alpha_{0}\,\omega_{a}+\alpha_{2}\sigma_{a}\right)+\alpha_{2}\delta\sigma^{a}\omega_{a}\right]\,.\label{var_I}
\end{equation}
In Riemannian geometry, $\omega^{a}$
is an on-shell function of $e^{a}$ and $\delta\omega^{a}$ can
be expressed in terms of $\delta e^{a}$, but we omit this step and work directly with $\delta\omega^a(e)$. An explicit form of $\delta I_{\mathrm{on-shell}}$ depends on the location
of the boundary $\partial\mathcal{M}$.

Throughout this work, it is convenient to deal with different representations of the
Minkowski metric. We adopt the following notation: In
the light-cone representation, the Minkowski metric is
\begin{equation}
\eta_{ab}=\begin{pmatrix}0 & 1 & 0\\
1 & 0 & 0\\
0 & 0 & 1
\end{pmatrix}\,,\label{ndm}
\end{equation}
whereas the diagonal Minkowski metric is denoted by $\bar{\eta}=$diag$(-1,1,1)$.
Since gauge fields are constructed with respect to a given tangent
space metric, in the diagonal case they must also be distinguished
with bars, namely $\bar{e}$, $\bar{\omega}$ and $\bar{\sigma}$.
The Levi-Civita tensor is normalized as $\epsilon_{012}=1$. We label in the same way both sets of Lorentz indices $a,b,\ldots$
because they cannot be confused.

\subsection{Stationary solutions}

In this section we focus on stationary solutions, which correspond to manifolds of the form $\mathcal{M}=\mathbb{R}\times\Sigma$,
with $\mathbb{R}$ the time-line and $\Sigma$ a spatial section.
The local coordinates are $x^{\mu}=(t,r,\varphi)$ and $x^{i}=(t,\varphi)$
are the boundary coordinates, with the boundary $\partial\mathcal{M}=\mathbb{R}\times\partial\Sigma$
placed at the radial infinity. The polar angle parametrizes $\partial\Sigma$.

The general solution of the equations of motion in the Gauss normal
frame and the ADM form of the metric reads
\begin{equation}
ds^{2}=-N^{2}dt^{2}+\frac{dr^{2}}{N^{2}}+r^{2}\left(d\varphi+N_{\varphi}dt\right)^{2}\,,\label{btz1}
\end{equation}
where
\begin{equation}
N^{2}=-M+\frac{J^{2}}{4r^{2}}\,,\qquad N_{\varphi}=-\frac{J}{2r^{2}}\,,\label{btz6-1}
\end{equation}
and $M,J$ are integration constants. The vielbein can be chosen as
\begin{align}
\bar{e}^{0} & =Ndt\,,\nonumber \\
\bar{e}^{1} & =N^{-1}dr\,,\label{btz3}\\
\bar{e}^{2} & =r\left(d\varphi+N_{\varphi}dt\right)\,,\nonumber
\end{align}
and the field equations determine completely the torsionless spin-connection
in terms of $M$ and $J$, yielding
\begin{align}
\bar{\omega}^{0} & =Nd\varphi\,,\nonumber \\
\bar{\omega}^{1} & =\frac{J}{2r^{2}N}\,dr\,,\label{btz3-1}\\
\bar{\omega}^{2} & =-\frac{J}{2r}\,d\varphi\,.\nonumber
\end{align}
The gravitational Maxwell field depends on two additional arbitrary constants $c$ and $b$,
\begin{align}
\bar{\sigma}^{0}& =2c\,Ndt+2r\,\left( W^{\prime }+NX\right) dr-\frac{JS}{r}\,d\varphi \,,  \notag \\
\bar{\sigma}^{1}& =2\left( S'+\frac{b}{N}+\frac{3r^{2}}{2JN}\right) dr-JWd\varphi \,,\label{btz8} \\
\bar{\sigma}^{2}& =-\left( \frac{Jc}{r}-r\right) dt-JXdr+2\,\left(rb+NS+\frac{r^{3}}{2J}\right) d\varphi \,,  \notag
\end{align}
and three arbitrary functions $S(r)$, $W(r)$ and $X(r)$. This solution was first reported in \cite{Hoseinzadeh:2014bla} and it exists only when $\alpha_2 \neq 0$. From eq.(\ref{cs5-1}), it seems that setting $\alpha_2=0$ leaves $\sigma_a$ unconstrained, but in that case $\sigma_a$ vanishes from the action as a fundamental field, meaning that its contribution to the dynamics is identically zero. When $\alpha_2 \neq 0$, on the other hand, it is always $\sigma_a \neq 0$. Two cases ($\alpha_2 = 0$ and $\alpha_2 \neq 0$),  therefore, describe different branches of the field equations. From this point of view, it is clear that the branch described by eq.(\ref{cs5'-1}) and the solution we found in (\ref{btz8}) correspond to $\alpha_2 \neq 0$. An observation that the $\alpha_2 = 0$ limit is not well-defined at the level of the solution is supported by the fact that there is no choice of the arbitrary functions that can continuously switch off this field and move to the  $\alpha_2 = 0$ case. However, as we will see below, the conserved charges have this limit ($\alpha_2 =0$) well-defined.

Let us require that the above solution has a well-defined non-singular $J\rightarrow 0$ limit in the space of parameters, describing a non-rotating geometry when $\alpha_2 \neq 0$. This will give us an insight about a choice of asymptotic conditions for the arbitrary functions. Indeed, the regularity of the arbitrary functions in \eqref{btz8} constrains them to behave for $J\rightarrow 0$ as
\begin{equation}
N =  \mathcal{O}_{J}(1)\,, \qquad  X,W =  \mathcal{O}_{J}(1/J)\,, \qquad
S =  -\frac{r^{3}}{2JN}+\mathcal{O}_{J}(1)\,,  \label{J=0 behavior}
\end{equation}
and, in addition, $W'+NX=\mathcal{O}_{J}(1)$ must be satisfied. Note that $N'=\mathcal{O}_{J}(J^2)$. Since the gravitational part of the solution is completely gauge fixed, it is natural also to fix the extra gauge field and remove the arbitrary functions, because they are not physical. We choose to fix it as
\begin{equation}
S=-\frac{1}{N}\,\left(\frac{r^{3}}{2J}+br\right)\,,\qquad W=\frac{S}{r}\,,\qquad X=\frac{r}{JN^{2}}\,,\label{app8-1}
\end{equation}
because, as we will see later, in that case this stationary solution can be obtained as a particular point of the non-stationary solution discussed in the next section.

The flat (Minkowski) space,  when $\alpha_2\neq0$, corresponds to the point $(M,J,b,c)=(-1,0,b,c)$. Interestingly, the gravitational Maxwell field with the charges $b,c\neq 0$ does not curve the Minkowski geometry, but it contributes to the conserved charges. For instance, the $J=0$ sector describes the static metric manifold, but the total angular momentum of the system, which is the conserved charge associated to a rotational Killing vector $\partial_{\varphi}$, can be non-vanishing. We will calculate conserved quantities below.

The geometry of the solution (\ref{btz1}) depends on the range of the parameters. We are interested in flat cosmologies with $M>0$ and $J\in \mathbb{R}$,
whose entropy and thermodynamics in GR were discussed in Ref.~\cite{Barnich:2012xq}, and charges and vacuum energy in Ref.~\cite{Miskovic:2016mvs}.

In order to check that the action principle is satisfied, namely, that the variation (\ref{var_I}) vanishes on-shell for suitable boundary conditions, we set the boundary at $r=const,$ and use the identification $\epsilon^{ij}\equiv\epsilon^{irj}$, which implies
$\epsilon^{t\varphi}=-1$. We find
\begin{eqnarray}
\delta I_{\mathrm{on-shell}} & = & -\frac{k}{4\pi}\int\limits_{\partial \mathcal{M}} dtd\varphi\,\left[-\delta\bar{e}_{t}^{0}\bar{\omega}_{\varphi}^{0}+\delta\bar{e}_{t}^{2}\left(\bar{\omega}_{\varphi}^{2}+\alpha_{2}\bar{e}_{\varphi}^{2}\right)-\alpha_{2}\delta\bar{e}_{\varphi}^{2}\bar{e}_{t}^{2}\right.\nonumber \\
 &  & +\left.\delta\bar{\omega}_{\varphi}^{0}\left(\bar{e}_{t}^{0}+\alpha_{2}\bar{\sigma}_{t}^{0}\right)-\delta\bar{\omega}_{\varphi}^{2}\left(\bar{e}_{t}^{2}+\alpha_{2}\bar{\sigma}_{t}^{2}\right)-\alpha_{2}\delta\bar{\sigma}_{t}^{0}\bar{\omega}_{\varphi}^{0}+\alpha_{2}\delta\bar{\sigma}_{t}^{2}\bar{\omega}_{\varphi}^{2}\right]\,.
\end{eqnarray}
Based on the solution (\ref{btz3}-\ref{btz8}), the following on-shell relations apply for the spin-connection $\bar{\omega}_i^a(\bar{e})$, which is not an independent field,
\begin{equation}
\bar{\omega}_{\varphi}^{0}=\bar{e}_{t}^{0}\,,\qquad \bar{\omega}_{\varphi}^{2}=\bar{e}_{t}^{2}\,.
\end{equation}
After this condition has been applied, $\delta I_{\mathrm{on-shell}}$ only depends on $\delta\bar{e}_i^a$ and $\delta\bar{\sigma}_a^i$, and not on the variation of the extrinsic curvature of the boundary. Furthermore, the remaining terms are coming only from the gravitational Maxwell contributions in the action because they are proportional to the Maxwell coupling constant $\alpha_2$. Imposing the following boundary conditions on the gravitational Maxwell field, fulfilled for the stationary solutions (\ref{btz3}-\ref{btz8}), and  defining the fall-off
\begin{equation}
\begin{array}{ll}
\bar{e}_{t}^{0}=\mathcal{O}(1)\,,\qquad & \bar{\sigma}_{t}^{0}=\beta\,\bar{e}_{t}^{0}=\mathcal{O}(1)\,,\\
\bar{e}_{t}^{2}=\mathcal{O}(1/r)\,, & \bar{\sigma}_{t}^{2}=\bar{e}_{\varphi}^{2}+\mathcal{O}(1/r)\,,
\end{array}
\end{equation}
where $\beta$ is some proportionality constant ($\beta=2c$), we find that the on-shell variation of the action identically vanishes,
\begin{equation}\label{dIos}
\delta I_{\mathrm{on-shell}}=0\,.
\end{equation}
We conclude that the action principle is satisfied without addition of a boundary term, provided the leading order in the fall-off of the vielbein is kept fixed on the boundary. This has two consequences. First, the variation of the action is finite, meaning that the CS formulation of the action for the Maxwell group does not need additional counterterms. Second, it requires the leading order of the vielbein to be fixed on the boundary, which means $M$ and $J$. These boundary conditions are not the Dirichlet ones for the induced boundary fields. This is similar to a field theory approach to three-dimensional GR,
whose dynamics is given by the CS action without boundary terms in both asymptotically flat space \cite{Miskovic:2016mvs,Detournay:2014fva}
and asymptotically AdS space \cite{Banados:1998ys,Miskovic:2006tm}.
While in the flat case this provides a well-defined variational principle and
well-defined one-point functions \cite{Detournay:2014fva}, in the GR AdS case this is motivated
by the IR finiteness of the theory and the well-definiteness of the action principle. In the latter one, the boundary conditions are the Dirichlet ones for the leading order $g_{(0)ij}(x)$ in the fall-off of the induced metric $h_{ij}(r,x)=r^2\,g_{(0)ij}(x)+{\cal O}(1)$. Thus, the induced metric itself does not satisfy the Dirichlet boundary conditions \cite{Miskovic:2006tm}.

Now we are ready to compute conserved charges. In
CS gravity, Noether charges associated to the asymptotic
Killing vectors $\xi=\xi^{i}\partial_i$ are evaluated at the asymptotic infinity and at a constant
time slice $\partial\Sigma$. They have the form~\cite{Julia:1998ys,Feng:1999mk}
\begin{equation}\label{qdiff}
Q[\xi]=\frac{k}{4\pi}\,\int\limits _{\partial\Sigma}\langle A\iota_{\xi}A\rangle\,.
\end{equation}
In particular,
for the gauge field (\ref{cs1}), they become
\begin{equation}
Q[\xi]=\frac{k}{4\pi}\lim_{r\rightarrow \infty }\int\limits_{0}^{2\pi} d\varphi \, \xi^{i}\left[\alpha_{0}\bar{\omega}_{\varphi}^{a}\bar{\omega}_{ai}+\bar{e}_{\varphi}^{a}\bar{\omega}_{ai}+\bar{\omega}_{\varphi}^{a}\bar{e}_{ai}+\alpha_{2}\left(\bar{e}_{\varphi}^{a}\bar{e}_{ai}+\bar{\omega}_{\varphi}^{a}\bar{\sigma}_{ai}+\bar{\sigma}_{\varphi}^{a}\bar{\omega}_{ai}\right)\right]\,.\label{eq:charge-1}
\end{equation}
Using this formula, we calculate the mass $m$ and the angular momentum
$j$ of the solution (\ref{btz3}-\ref{btz8}) for time translations
$\xi=\partial_{t}$ and rotations $\xi=\partial_{\varphi}$,
as
\begin{eqnarray}
m & \equiv & Q[\partial_{t}]=\frac{k}{2}\left[M+\alpha_{2}\left(2Mc-\frac{J}{2}\right)\right]\,,\nonumber \\
j & \equiv & Q[\partial_{\varphi}]=\frac{k}{2}\left(-\frac{J}{2}+\alpha_{0}M-2\alpha_{2}bJ\right)\,.\label{mj}
\end{eqnarray}
Remarkably, the gravitational Maxwell field contributes to the total mass and angular
momentum of the solution and therefore modifies the asymptotic sector.
As we have set $\alpha_{1}=1$, the above charges match the EH
ones, with and without the gravitational CS term \cite{Banados:1994tn}, when the gravitational Maxwell
interaction vanishes. Even though the calculations for $\alpha_2 =0$ and $\alpha_2\neq0$ have to be done independently, the final charge formula (\ref{mj}) holds in both cases. In the total static limit ($j=0$), the manifold has to rotate with the momentum $J=\frac{2\alpha_{0}M}{1+4\alpha_{2}b}$ in order to compensate the rotation of the gravitational Maxwell field, and the mass becomes
\begin{equation}
m_{\mathrm{static}}=\frac{k M}{2}\left(1+2c\alpha_2-\frac{\alpha_{0}\alpha_2}{1+4\alpha_2b}\right)\,.
\end{equation}
On the other hand, the vacuum $J=0$
and $M=-1$ has the Minkowski spacetime geometry with the total energy and angular momentum
\begin{equation}
m_{0}=-\frac{k }{2}\,\left( 1+2c\alpha _{2}\right) \,,\qquad j_{0}=-%
\frac{k \alpha _{0}}{2}\,.
\end{equation}%
Thus, the vacuum energy $m_{0}$ depends on the charge $c$ and suitably charged
gravitational Maxwell field can make it vanish, unlike in GR where always $m_{0} \neq 0$
\cite{Barnich:2012aw,Miskovic:2016mvs}. The total angular momentum of the vacuum, $j_{0}$, never
vanishes.

Let us remember that our goal is to find an asymptotic algebra associated to the Maxwell algebra. This cannot be done by looking at the stationary solutions, because their asymptotic symmetries at the time-like boundary are trivial. For example, the $\mathfrak{bms}_3$ algebra can be obtained as asymptotic symmetry at spatial boundary of asymptotically flat Eistein gravity when a more
general set of non-stationary solutions is considered \cite{Compere:2017knf}. We shall, however, turn our attention to spacetimes with null boundary, where non-stationary solutions in three-dimensional gravity with Maxwell symmetry are known to exist.

\subsection{Solution with null boundary }

Space-times with null boundary are better described in the BMS gauge, where the manifold is parameterized by the local coordinates $x^{\mu}=(u,r,\phi)$, with $u$ being the retarded time coordinate and the boundary is located at $r=const\rightarrow\infty$. Then, using diffeomorphisms, the metric can be cast in the form
\begin{equation}
ds^{2}=\mathcal{M}du^{2}-2dudr+\mathcal{N}d\phi du+r^{2}d\phi^{2}\,.\label{af9}
\end{equation}
Furthermore, since the solutions of Einstein equations are also solutions of the CS gravity theory invariant under the Maxwell group that we are studying, we can immediately write down the well-known results for the metric in asymptotically flat three-dimensional gravity \cite{Barnich:2013yka},
\begin{equation}
{\cal M={\cal M{\rm (\phi),\qquad}}}\mathcal{N}=\mathcal{J}(\phi)+u\mathcal{M^{\prime}}(\phi)\,.\label{eq:af8}
\end{equation}
By using light-cone coordinates in tangent space, the line element can be written in terms of vielbein one-forms as
\begin{equation}
ds^{2}=2e^{0}e^{1}+\left(e^{2}\right)^{2}\,,\label{af11}
\end{equation}
where
\begin{align}
e^{0} & =-dr+\frac{\mathcal{M}}{2}\,du+\frac{\mathcal{N}}{2}\,d\phi\,,\nonumber \\
e^{1} & =du\,,\label{af13}\\
e^{2} & =rd\phi\,.\nonumber
\end{align}
This leads to the following Levi-Civita spin connection
\begin{align}
\omega^{0} & =\frac{\mathcal{M}}{2}\,d\phi\,,\nonumber \\
\omega^{1} & =d\phi\,,\label{af19}\\
\omega^{2} & =0\,.\nonumber
\end{align}

Now we have to solve the gravitational Maxwell field $\sigma^{a}$ from the last equation in (\ref{cs5'-1}). As it is well-known \cite{Banados:1994tn}, the
component $A_{r}$ of the gauge field of the action in the radial foliation is a Lagrange multiplier, and it is standard to gauge-fix
it as $\partial_{\phi}A_{r}=0$. The gravitational fields (\ref{af13})
and (\ref{af19}) fulfil this condition. Applied to the $\sigma^{a}$ field,
one gets $\sigma_{r}^{a}=\sigma_{r}^{a}(u,r)$. We choose the simplest gauge-fixing,
\begin{equation}
\sigma_r^a =0\,.\label{bms8}
\end{equation}
In this ansatz, the general solution becomes
\begin{align}
\sigma^{0} & =\frac{1}{2}\left(\mathcal{F}-r^{2}\right)d\phi+\left(\frac{1}{2}\mathcal{\,M}\mathcal{Q}-\mathcal{\dot{I}}+\mathcal{P}^{\prime}+\frac{1}{2}\mathcal{\,N}\right)du\,,\nonumber \\
\sigma^{1} & =\mathcal{Q}\, du+\mathcal{B}\,d\phi\,,\label{bms11}\\
\sigma^{2} & =\left(r+\mathcal{\dot{B}}-\mathcal{Q}^{\prime}\right)du+\mathcal{I}\,d\phi\,,\nonumber
\end{align}
where the functions of the boundary coordinates,  $\mathcal{Q}(u,\phi)$, $\mathcal{F}(u,\phi)$,
${\cal P}(u,\phi)$, $\mathcal{B}\left(u,\phi\right)$ and $\mathcal{I}(u,\phi)$,
are subjected to the differential equation
\begin{equation}
\mathcal{\dot{F}}+2\mathcal{\dot{I}}^{\prime}+{\cal M}{\cal B}-2\mathcal{P}^{\prime\prime}-\mathcal{J}^{\prime}-u\mathcal{M}^{\prime\prime}-\mathcal{M}^{\prime}\mathcal{Q}-\mathcal{{\rm 2}M}\mathcal{Q}^{\prime}=0\,.\label{const}
\end{equation}
Here primes and dots stand for derivatives with respect to the coordinates
$\phi$ and $u$, respectively. The solution (\ref{bms11}) for the gravitational Maxwell field was not reported in the literature before. The existence of arbitrary functions of
the boundary coordinates $u$ and $\phi$ suggests a large number
of asymptotic symmetries. In particular, this solution matches the
stationary one when the functions ${\cal M}$ and ${\cal N}$
become constant,
\begin{equation}
\mathcal{M}(\phi)=M\,,\qquad\mathcal{N}(u,\phi)=-J\,.\label{bms14}
\end{equation}
Indeed, the stationary and null boundary solutions
are equivalent in the particular point (\ref{bms14}), as it can be seen by performing a coordinate transformation in a similar way as in Ref.~\cite{Barnich:2012aw}. Redefining $t=u+f\left(r\right)\,$, and $\varphi=\phi+g\left(r\right)$
with $f^{\prime}=N^{-2}$ and $g^{\prime}=-N_{\varphi}N^{-2}$, the
stationary metric in the ADM form (\ref{btz1}) becomes
\begin{equation}
ds^{2}=\left(r^{2}N_{\varphi}^{2}-N^{2}\right)du^{2}-2dudr+2r^{2}N_{\varphi}d\phi du+r^{2}d\phi^{2}\,.\label{app_2-1}
\end{equation}
We use the identity $\bar{\eta}_{ab}\bar{e}^{a}\bar{e}^{b}=\eta_{ab}e^{a}e^{b}$ to change the diagonal to light-cone Lorentz coordinates.
The vielbeins are related by
\begin{equation}
e_{\mu}^{a}=K_{\ b}^{a}(x)\bar{e}_{\mu}^{b}\,.\label{app_5-1-1}
\end{equation}
The same transformation is valid for the gravitational Maxwell field in the BMS gauge, $\sigma_{\mu}^{a}=K_{\ b}^{a}\bar{\sigma}_{\mu}^{b}$.
Note that $K_{\ b}^{a}$ is not a Lorentz tensor of rank two because
the upper and lower indices correspond to different coordinate frames. In the above
formula, the vielbein $\bar{e}_{\mu}^{b}$ can be deduced from equation
(\ref{app_2-1}) as
\begin{equation}
\bar{e}^{0}=Ndu+N^{-1}dr\,,\qquad \bar{e}^{1}=N^{-1}dr\,,\qquad\bar{e}^{2}=r\left(d\phi+N_{\varphi}du\right)\,,\label{app_3-1}
\end{equation}
while the vielbein $e_{\mu}^{a}$ is the one written in the BMS gauge
given by eq.(\ref{af13}). We can also use the identity $r^{2}N_{\varphi}^{2}-N^{2}={\cal M}=M$
valid in the considered particular point to obtain
\begin{equation}
K_{\ b}^{a}=e_{\mu}^{a}\bar{e}_{b}^{\mu}=\left(\begin{array}{ccc}
-\frac{1}{2N}\,\left(N_{\varphi}^{2}r^{2}+N^{2}\right) & \frac{1}{2N}\,\left(N_{\varphi}^{2}r^{2}-N^{2}\right) & rN_{\varphi}\\
N^{-1} & -N^{-1} & 0\\
-rN_{\varphi}N^{-1} & rN_{\varphi}N^{-1} & 1
\end{array}\right)\,.\label{app6-1}
\end{equation}
The knowledge of $K_{\ b}^{a}$ is sufficient to obtain all functions in the new reference frame.

On the other hand, a large number of arbitrary functions in \eqref{bms11} indicates a rich asymptotic structure of this topological theory. We would like to show that they induce an enlargement of the asymptotic BMS algebra in three-dimensional GR. To this end, and to obtain an explicit realization  of this enlargement, we will set to zero as many arbitrary functions as possible,  say $\mathcal{I}$, $\mathcal{P}$, $\mathcal{B}$, $\mathcal{Q}$, so that the gravitational
Maxwell field reduces to
\begin{align}
\sigma^{0} & =\frac{1}{2}\left(\mathcal{F}-r^{2}\right)\,d\phi+\frac{1}{2}\mathcal{\,N}du\,,\nonumber\\
\sigma^{1}& =0\,,\label{af20} \\
\sigma^{2}& =rdu\,.\nonumber
\end{align}
This leaves only one function as dynamical, ${\cal F}$, which can be solved from the constraint (\ref{const}) and gives
\begin{eqnarray}
\mathcal{F} & = & {\cal Z}+u{\cal J}^{\prime}+\frac{u^{2}}{2}{\cal M}^{\prime\prime}\,.\label{eq:efe}
\end{eqnarray}
Note that here still remains one arbitrary function ${\cal Z}(\phi)$ in the expression for the gravitational Maxwell field (\ref{af20}). At the particular point (\ref{bms14}) where this solution reduces to the stationary one, all the remaining
functions become constant, that is
\begin{equation}
{\cal M}(\phi)=M\,,\qquad {\cal J}(\phi)=-J\,,\qquad {\cal Z}(\phi)=-2Jb\,.\label{MJZ}
\end{equation}
At the same time, the functions $X(r)$, $W(r)$ and $S(r)$ appearing in eq.$\,(\ref{btz8})$ have to be gauge fixed as shown in eq.(\ref{app8-1}).
We conclude that the solution in the BMS gauge contain the stationary solutions described in the previous subsection, as they reduce to them when (\ref{MJZ}) is satisfied. The above gauge fixing is consistent with the one obtained in eq. (\ref{J=0 behavior}).

Finally, we have to ensure that the action principle is well-defined when geometries with null boundary are considered. We will show in the next section that, in CS theories, the radial dependence can be dropped out from the gauge field $A$ through the gauge transformation
\begin{equation}\label{gtransformation}
A=h^{-1}dh+h^{-1}ah\,,
\end{equation}
where $h=$e$^{-rP_{0}}$. Then, the asymptotic field $a$ is $r$-independent and the on-shell action (\ref{Var}) becomes
\begin{equation}
\delta I_{\mathrm{on-shell}}=\frac{k}{4\pi }\int\limits_{\partial \mathcal{M}}\left\langle \delta \left( h^{-1}dh+h^{-1}ah\right) \left(h^{-1}dh+h^{-1}ah\right) \right\rangle \,.
\end{equation}
Since we are on the boundary $r=const\rightarrow\infty $, $h$ is constant and therefore $\delta h=0=dh$, which leads to
\begin{equation}
\delta I_{\mathrm{on-shell}}=\frac{k}{4\pi}\int\limits_{\partial
\mathcal{M}}\left\langle \delta aa\right\rangle \,.
\end{equation}
On the other hand, from eqs.(\ref{af13}), (\ref{af19}) and (\ref{af20}) we see that our solution satisfies the following boundary conditions for the fields $\omega^{a}$ and $\sigma^{a}$,
\begin{equation}
\omega _{\phi }^{a} =e_{u}^{a}\,,\qquad \omega_{u}^{a}=0\,,\qquad \sigma _{u}^{a} =e_{\phi }^{a}\,.
\end{equation}
Note that the first two conditions (on $\omega ^{a}$ and $e^{a}$) are the same as in the Poincar\'{e} case \cite{Barnich:2013yka}. Setting $\epsilon ^{u\phi }=1$ and $\omega _{u}^{a}=0$, the variation of the action reads
\begin{eqnarray}
\delta I_{\mathrm{on-shell}} &=&\frac{k}{4\pi }\int\limits_{\partial
\mathcal{M}}dud\phi \,\left[ \delta e_{u}^{a}\,\left( \omega _{a\phi
}+\alpha _{2}e_{a\phi }\right) -\alpha _{2}\,\delta e_{\phi
}^{a}e_{au}\right.   \nonumber \\
&&\left. -\delta \omega _{\phi }^{a}\,\left( e_{au}+\alpha _{2}\sigma
_{au}\right) +\alpha _{2}\delta \sigma _{u}^{a}\omega _{a\phi }\right] \,.
\end{eqnarray}
Now we apply $\omega _{\phi }^{a}=e_{u}^{a}$ and get
\begin{equation}
\delta I_{\mathrm{on-shell}}=\frac{k \alpha _{2}}{4\pi}
\int\limits_{\partial \mathcal{M}}dud\phi \,\left[ \delta e_{u}^{a}\left(
e_{a\phi}-\sigma_{au}\right) +\left(\delta\sigma_{u}^{a}-\delta e_{\phi}^{a}\right) e_{ua}\right] \,.
\end{equation}
So far, we have imposed only the boundary conditions on the fields $\omega _{i}^{a}$, which ensures that the $\left( e,\omega \right) $ sector of the action satisfies the variational principle. Now we have to require the last condition on the Maxwell gravitational field, $\sigma _{u}^{a}=e_{\phi }^{a}$, which leads to $\delta I_{\mathrm{on-shell}}=0$ for any value of the constant $\alpha _{2}$. Thus, in this case, the action principle is satisfied for any value of the coupling constants.

\newsection{Asymptotic symmetries}\label{three}
In the previous section, we found the following form of the gauge field in the BMS gauge,
\begin{eqnarray}
A & = & \left(-dr+\frac{1}{2}{\cal \,M}du+\frac{1}{2}\mathcal{\,N}d\phi\right)\,P_{0}+du\,P_{1}+rd\phi\,P_{2}+\frac{1}{2}\,\mathcal{M}d\text{\ensuremath{\phi}\,}J_{0}+d\phi\,J_{1}\nonumber \\
 &  & +\frac{1}{2}\left(\mathcal{N}du+\mathcal{F}d\phi-r^{2}d\phi\right)\,Z_{0}+rdu\,Z_{2}\,.\label{eq:AMaxwell}
\end{eqnarray}
For the moment, let us assume that the functions ${\cal M}$ and ${\cal N}$ and $\cal{F}$ (and therefore ${\cal J}$ and ${\cal Z}$) depend on all boundary coordinates $x^{i}=(u,\phi)$, and we shall set them on-shell later.
The radial dependence of the gauge field $A$ can be gauged away by
a gauge transformation \eqref{gtransformation}. Using the identity $h^{-1}dh=-drP_{0}$ and the Baker-Campbell-Hausdorff formula, we obtain
\begin{equation}
h^{-1}ah=a+rdu\,Z_{2}+rd\phi\,P_{2}-r^{2}d\phi\,Z_{0}\,.
\end{equation}
The final effect is that the radial dependence from the gauge field $A$
is dropped out and the new gauge field $a$ becomes the asymptotic field,
\begin{equation}
a =  \frac{1}{2}\left({\cal M}du+{\cal N}d\phi\right)\,P_{0}+du\,P_{1}+\frac{1}{2}{\cal \,M}d\text{\ensuremath{\phi}}\,J_{0}+d\phi\,J_{1}+\frac{1}{2}\left({\cal N}du+\mathcal{F}d\phi\right)\,Z_{0}\,.\label{eq:a}
\end{equation}

The asymptotic symmetry is a residual symmetry,
that leaves the asymptotic conditions (\ref{eq:AMaxwell}) invariant. In order to find it, we consider gauge parameters of the form
\begin{equation}
\Lambda=h^{-1}\lambda h \qquad,\qquad \lambda=\varepsilon^{a}(u,\phi)\,P_{a}+\chi^{a}(u,\phi)\,J_{a}+\gamma^{a}(u,\phi)\,Z_{a}\,.\label{eq:lambda}
\end{equation}
Note that the gauge parameter components in \eqref{eq:lambda} are not the same as the ones introduced in \eqref{cs7}, as in this case they only depend on the boundary coordinates. Then, gauge transformations of the full connection $A$ with gauge parameter $\Lambda$ lead to $r$-independent gauge transformations of $a$ with gauge parameter $\lambda$. Now, we require that the transformed field, $a+D\lambda$, and the original one,
$a$, have the same form (\ref{eq:a}). A change of the boundary field (\ref{eq:a}) is given by
\begin{eqnarray}
\delta_{\lambda}a&=&\frac{1}{2}\left[\,\text{\ensuremath{\delta_{\lambda}}}{\cal M}(u,\phi)\, du+\frac{1}{2}\text{\,\ensuremath{\delta_{\lambda}}}{\cal N}(u,\phi)\,d\phi\right]P_{0}+\frac{1}{2}\,\text{\ensuremath{\delta_{\lambda}}}{\cal M}(u,\phi)\, d \phi\, J_{0}\nonumber \\  & &+\frac{1}{2}\left(\delta_{\lambda}{\cal N}(u,\phi)\,du+\delta_{\lambda}\mathcal{F}(u,\phi)\,d\phi\right)\,Z_{0}\,.\label{eq:aphi}
\end{eqnarray}
On the other hand, this change has to be equal to the gauge transformation $\delta_{\lambda}a=D\lambda$.
Replacing (\ref{eq:a}) and (\ref{eq:lambda}) in $D\lambda$, we
find the angular component of $a$ to be
\begin{eqnarray}
\delta_{\lambda}a_{\phi} & = & \left(\varepsilon^{0\prime}-\frac{{\cal N}}{2}\,\chi^{2}-\frac{{\cal M}}{2}\,\varepsilon^{2}\right)\,P_{0}+\left(\varepsilon^{1\prime}+\varepsilon^{2}\right)\,P_{1}+\left(\varepsilon^{2\prime}+\frac{{\cal N}}{2}\,\chi^{1}+\frac{{\cal M}}{2}\,\xi^{1}-\xi^{0}\right)\,P_{2}\nonumber \\
 &  & +\left(\chi^{0\prime}-\frac{{\cal M}}{2}\,\chi^{2}\right)\,J_{0}+\left(\chi^{1\prime}+\chi^{2}\right)\,J_{1}+\left(\text{\ensuremath{\chi}}^{2\prime}+\frac{{\cal M}}{2}\,\chi^{1}-\chi^{0}\right)\,J_{2}\label{deltaaphi}\\
 &  & +\left(\gamma^{0\prime}-\frac{{\cal F}}{2}\,\chi^{2}-\frac{{\cal N}}{2}\,\varepsilon^{2}-\frac{{\cal M}}{2}\,\gamma^{2}\right)\,Z_{0}+\left(\gamma^{1\prime}+\gamma^{2}\right)\,Z_{1}\nonumber\\& &
 +\left(\gamma^{2\prime}+{\cal \frac{F}{{\rm 2}}\,}\chi^{1}+\frac{{\cal N}}{2}\,\varepsilon^{1}+\frac{{\cal M}}{2}\,\gamma^{1}-\gamma^{0}\right)\,Z_{2}\,.\nonumber
\end{eqnarray}
Comparing this expression with the component along $\phi$ of (\ref{eq:aphi}),
we obtain the transformation law of the arbitrary functions on the boundary,
\begin{eqnarray}
\text{\ensuremath{\delta_{\lambda}}}{\cal M} & = & 2\chi^{0\prime}-{\cal M}\chi^{2}\,,\nonumber \\
\text{\ensuremath{\delta_{\lambda}}}{\cal N} & = & 2\varepsilon^{0\prime}-{\cal N}\chi^{2}-{\cal M}\varepsilon^{2}\,,\label{deltaMNF}\\
\delta_{\lambda}\mathcal{F} & = & 2\gamma^{0\prime}-\mathcal{F}\chi^{2}-{\cal N}\varepsilon-{\cal M}\gamma^{2}\,,\nonumber
\end{eqnarray}
where the parameters satisfy differential equations
\begin{equation}
\begin{array}{ll}
\gamma^{1\prime}+\gamma^{2}  =0\,,\medskip \qquad & \gamma^{2\prime}+{\cal \frac{F}{{\rm 2}}\,}\chi^{1}+\frac{{\cal N}}{2}\,\varepsilon^{1}+\frac{{\cal M}}{2}\,\gamma^{1}-\gamma^{0} =0\,,\\
\varepsilon^{1\prime}+\varepsilon^{2}  =0\,,\medskip\qquad & \varepsilon^{2\prime}+\frac{{\cal N}}{2}\,\chi^{1}+\frac{{\cal M}}{2}\,\varepsilon^{1}-\varepsilon^{0}  =0\,,\\
\chi^{1\prime}+\chi^{2} =0\,,\medskip\qquad\qquad & \text{\ensuremath{\chi}}^{2\prime}+\frac{{\cal M}}{2}\,\chi^{1}-\chi^{0}  =0\,.
\end{array}\label{aphi}
\end{equation}
Furthermore, the retarded time component of the gauge field $a$ transforms as
\begin{eqnarray*}
\delta_{\lambda}a_{u} & = & \left(\dot{\varepsilon}^{0}-\frac{{\cal M}}{2}\,\chi^{2}\right)\,P_{0}+\left(\dot{\varepsilon}^{1}+\chi^{2}\right)\,P_{1}+\left(\dot{\varepsilon}^{2}+\frac{{\cal M}}{2}\,\chi^{1}-\chi^{0}\right)\,P_{2}+\dot{\chi}^{a}\,J_{a}\\
 &  & +\left(\dot{\gamma}^{0}-\frac{\text{\ensuremath{{\cal N}}}}{2}\,\chi^{2}-\frac{{\cal M}}{2}\,\varepsilon^{2}\right)\,Z_{0}+\left(\dot{\gamma}^{1}+\varepsilon^{2}\right)\,Z_{1}+\left(\dot{\gamma}^{2}+\frac{\text{\ensuremath{{\cal N}}}}{2}\,\chi^{1}+\frac{{\cal M}}{2}\,\varepsilon^{1}-\text{\ensuremath{\varepsilon}}^{0}\right)\,Z_{2}\,.
\end{eqnarray*}
Comparing this expression with the $u$-component of (\ref{eq:aphi}),
we get
\begin{eqnarray}
\text{\ensuremath{\delta_{\lambda}}}{\cal M} & = & 2\dot{\varepsilon}^{0}-{\cal M}\chi^{2}\,,\medskip\nonumber \\
\text{\ensuremath{\delta_{\lambda}}}{\cal N} & = & 2\dot{\gamma}^{0}-{\cal N}\chi^{2}-{\cal M}\varepsilon^{2}\,,\label{deltaMN}
\end{eqnarray}
where $\dot{\chi}^{a}=0$, and the other parameters satisfy
\begin{equation}
\begin{array}{ll}
\dot{\gamma}^{1}+\varepsilon^{2}  =0\,,\medskip\qquad & \dot{\gamma}^{2}+\frac{\text{\ensuremath{{\cal N}}}}{2}\,\chi^{1}+\frac{{\cal M}}{2}\,\varepsilon^{1}-\text{\ensuremath{\varepsilon}}^{0}  =0\,,\\
\dot{\varepsilon}^{1}+\chi^{2}  =0\,,\medskip\qquad\qquad & \dot{\varepsilon}^{2}+\frac{{\cal M}}{2}\,\chi^{1}-\chi^{0}  =0\,.
\end{array}\label{au}
\end{equation}
Solving the equations (\ref{eq:aphi}) and (\ref{au}),
we arrive to the following solution,
\begin{equation}
\begin{array}{llllll}
\chi^{0} & =\frac{\mathcal{M}}{2}\,Y-Y^{\prime\prime}\,,\medskip & \varepsilon^{0} & =\frac{1}{2}\left(\mathcal{M}\varepsilon^{1}+\mathcal{N}Y\right)-\varepsilon^{1\prime\prime}\,, & \gamma^{0} & =\frac{1}{2}\left(\mathcal{M}\gamma^{1}+\mathcal{N}\varepsilon^{1}+\mathcal{F}Y\right)-\gamma^{1\prime\prime},\\
\chi^{2} & =-Y^{\prime}\,,\medskip & \varepsilon^{2} & =-T^{\prime}-uY^{\prime\prime}\,, & \gamma^{2} & =-R^{\prime}-uT{}^{\prime\prime}-\frac{u^{2}}{2}\,Y^{\prime\prime\prime}\,,\\
\chi^{1} & =Y\,, & \varepsilon^{1} & =T+uY^{\prime}\,, & \gamma^{1} & =R+uT^{\prime}+\frac{u^{2}}{2}\,Y^{\prime\prime}\,,
\end{array}
\end{equation}
where $Y=Y(\phi)$, $T=T(\phi)$ and $R=R(\phi)$ are arbitrary functions defined on $\partial\Sigma$. Now we impose the field equations and the functions ${\cal M}$, ${\cal N}$ and ${\cal F}$ acquire the form given by eqs.(\ref{eq:af8}) and (\ref{eq:efe}). The final transformation law for the arbitrary functions of the asymptotic connection $a$ turns out to be
\begin{eqnarray}
\delta{\cal M} & = & {\cal M}^{\prime}Y+2{\cal M}Y^{\prime}-2Y{}^{\prime\prime\prime}\,,\nonumber \\
\delta{\cal J} & = & {\cal M}^{\prime}T+2{\cal M}T^{\prime}-2T{}^{\prime\prime\prime}+{\cal J}^{\prime}Y+2{\cal J}Y^{\prime}\,,\label{trans}\\
\delta{\cal Z} & = & {\cal M}^{\prime}R+2{\cal M}R^{\prime}-2R{}^{\prime\prime\prime}+{\cal J}^{\prime}T+2{\cal J}T^{\prime}+{\cal Z}^{\prime}Y+2{\cal Z}Y^{\prime}\,,\nonumber
\end{eqnarray}
which contain the information of asymptotic symmetries and their algebra.

Let us now compute the charge algebra of the theory. As discussed in Ref.~\cite{Banados:1994tn}, the algebra is spanned by the conserved charges
$Q[\Lambda]$, which, as mentioned before, are on-shell equivalent to diffeomorphism charges of the form \eqref{qdiff}
with $\Lambda=\iota_{\xi}A$. Furthermore, the charge algebra in representation of Poisson brackets can be obtained
using the Regge-Teitelboim method \cite{Regge:1974zd} directly from the transformation law
\begin{equation}\label{rt}
\delta_{\Lambda_{2}}Q[\Lambda_{1}]=\left\{ Q[\Lambda_{1}],Q[\Lambda_{2}]\right\} .
\end{equation}
On the other hand, the variation of the charge in CS theory is given
by \cite{Banados:1994tn}
\begin{equation}
\delta Q[\Lambda]=\frac{k}{2\pi}\int\limits _{\partial\Sigma}\left\langle \Lambda\delta A\right\rangle \,.
\end{equation}
After applying the gauge transformation \eqref{gtransformation} which introduces the asymptotic field (\ref{eq:a}), and using \eqref{eq:lambda}, we get
\begin{equation}
\delta Q[\lambda]=\frac{k}{2\pi}\int d\phi\left\langle \lambda\delta a_{\phi}\right\rangle \,.
\end{equation}
Since the invariant tensor is known,
as well as the gauge field $a$, after a straightforward calculation
one arrives to
\begin{equation}
\delta Q[Y,T,R]=\frac{k}{4\pi}\int d\phi\left[Y\left(\alpha_{2}\delta{\cal Z}+\alpha_{0}\delta{\cal M}+\delta{\cal J}\right)+T\left(\alpha_{2}\delta{\cal J}+\delta{\cal M}\right)+\alpha_{2}R\delta{\cal M}\right]\,.
\end{equation}
We assume that the functions $Y$, $T$ and $R$ do not depend on the fields, in which case
it is trivial to integrate the variation out, finding
\begin{equation}
Q[Y,T,R]=\frac{k}{4\pi}\int d\phi\left[Y\left(\alpha_{2}{\cal Z}+\alpha_{0}{\cal M}+{\cal J}\right)+T\left(\alpha_{2}{\cal J}+{\cal M}\right)+\alpha_{2}R{\cal M}\right]\,. \label{Q}
\end{equation}
Now we define the asymptotic charges which correspond to the independent terms in (\ref{Q}),
\begin{eqnarray}
j[Y] & = & \frac{k}{4\pi}\int d\phi\,Y\left(\alpha_{2}{\cal Z}+{\cal J}+\alpha_{0}{\cal M}\right)\,,\nonumber \\
p[T] & = & \frac{k}{4\pi}\int d\phi\,T\left(\alpha_{2}{\cal J}+{\cal M}\right)\,,\\
z[R] & = & \frac{k}{4\pi}\int d\phi\,\alpha_{2}R{\cal M}\,.\nonumber
\end{eqnarray}
Using \eqref{rt}, they give rise to the centrally
extended Poisson algebra
\begin{eqnarray}
\left\{ j[Y_{1}],j[Y_{2}]\right\}  & = & j\left[[Y_{1},Y_{2}]\right]-\frac{k\alpha_{0}}{2\pi}\int d\phi\,Y_{1}Y_{2}^{\prime\prime\prime}\,,\nonumber \\
\left\{ j[Y],p[T]\right\}  & = & p\left[[Y,T]\right]-\frac{k}{2\pi}\int d\phi\,YT^{\prime\prime\prime}\,,\nonumber \\
\left\{ j[Y],z[R]\right\}  & = & z\left[[Y,R]\right]-\frac{k\alpha_{2}}{2\pi}\int d\phi\,YR^{\prime\prime\prime}\,,\\
\left\{ p[T_{1}],p[T_{2}]\right\}  & = & z\left[[T_{1},T_{2}]\right]-\frac{k\alpha_{2}}{2\pi}\int d\phi\,T_{1}T_{2}^{\prime\prime\prime}\,,\nonumber \\
\left\{ p[T],z[R]\right\}  & = & 0\,,\nonumber \\
\left\{ z[R_{1}],z[R_{2}]\right\}  & = & 0\,,\nonumber
\end{eqnarray}
where here $[x,y]=xy^{\prime}-yx^{\prime}$ stands for the Lie bracket of the vector
field components $x(\phi)$ and $y\left(\phi\right)$ on $\partial\Sigma$.
The result describes a deformed $\mathfrak{bms}_{3}$ algebra, as
expected, which is an infinite-dimensional enhancement of the Maxwell
algebra \cite{Caroca:2017onr}, with three central charges. It is common to write down the algebra
in Fourier modes,
\begin{equation}
{\cal J}_{m}=j[e^{im\text{\ensuremath{\phi}}}]\text{\,,\qquad}{\cal P}_{m}=p[e^{im\text{\ensuremath{\phi}}}]\text{\,,\qquad}{\cal Z}_{m}=z[e^{im\text{\ensuremath{\phi}}}]\,,\qquad m\in\mathbb{Z}\,,
\end{equation}
with all $\phi$-dependent functions expanded
on the circle $\partial\Sigma$. The corresponding deformed $\mathfrak{bms}_{3}$ symmetry reads
\begin{equation}
\begin{array}{lcl}
i\left\{ \mathcal{J}_{m},\mathcal{J}_{n}\right\}  & = & \left(m-n\right)\mathcal{J}_{m+n}+\dfrac{c_{1}}{12}\,m^{3}\delta_{m+n,0}\,,\medskip\\[6pt]
i\left\{ \mathcal{J}_{m},\mathcal{P}_{n}\right\}  & = & \left(m-n\right)\mathcal{P}_{m+n}+\dfrac{c_{2}}{12}\,m^{3}\delta_{m+n,0}\,,\medskip\\[6pt]
i\left\{ \mathcal{P}_{m},\mathcal{P}_{n}\right\}  & = & \left(m-n\right)\mathcal{Z}_{m+n}+\dfrac{c_{3}}{12}\,m^{3}\delta_{m+n,0}\,,\medskip\\[6pt]
i\left\{ \mathcal{J}_{m},\mathcal{Z}_{n}\right\}  & = & \left(m-n\right)\mathcal{Z}_{m+n}+\dfrac{c_{3}}{12}\,m^{3}\delta_{m+n,0}\,,\medskip\\[6pt]
i\left\{ \mathcal{P}_{m},\mathcal{Z}_{n}\right\}  & = & 0\,,\medskip\\[6pt]
i\left\{ \mathcal{Z}_{m},\mathcal{Z}_{n}\right\}  & = & 0\,,\medskip
\end{array}\label{dbms3}
\end{equation}
where we have used the integral representation of the Kronecker delta $\delta_{mn}=\frac{1}{2\pi}\int d\phi\,e^{i(m-n)\text{\ensuremath{\phi}}}$.
In the above classical algebra, the constants $\frac{c_{i}}{12}\,m^{3}\delta _{m+n,0}$, appearing on the
r.h.s., cannot be removed by a redefinition of the generators $%
\mathcal{J}_{m}\rightarrow \mathcal{J}_{m}+c\,\delta _{m,0}$, where $c$ is some constant, and
similarly for other generators. Thus, they present a non-trivial central
extension of this algebra. The three central charges are associated to three
terms in the gravitational action: the standard one $c_{2}=12k$ along the EH
term, the gravitational CS one $c_{1}=12k\alpha_{0}$ and, finally,
a new ingredient, a charge along the gravitational Maxwell term with torsion,
$c_{3}=12k\alpha_{2}$.
Note that the Maxwell algebra is a finite subalgebra of (\ref{dbms3})
formed by the generators $\left\{ {\cal J_{{\rm -1}}{\rm ,}J_{{\rm 0}}{\rm ,{\cal J}}_{{\rm 1}}{\rm ,{\cal P}_{-1},{\cal P}_{0}}{\rm ,{\cal P}_{1}}{\rm ,{\cal Z}_{-1}}{\rm ,}{\cal Z}_{{\rm 0}}{\rm ,}{\cal Z}}_{1}\right\} $
(see \cite{Caroca:2017onr}).

\newsection{Discussion}\label{four}

In this paper we have studied conserved charges and asymptotic symmetries of three-dimensional Chern-Simons
gravity invariant under action of the Maxwell group. This asymptotically flat gravity is characterized by three coupling constants. The first two are familiar: $\alpha_1=1$  is the strength of Einstein interaction and $\alpha_0$  the coupling gravitational CS term. The third one is the constant $\alpha_2$ and determines the strength of the gravitational Maxwell field, which is always non-vanishing (even for Minkowski space). In consequence, compared to Einstein theory, the last field modifies both the vacuum of the theory and its asymptotic sector. Indeed, we showed that the total energy and angular momentum of the Minkowski vacuum for a family of stationary cosmologies realized in asymptotically flat spacetimes with time-like boundary depend on $\alpha_2$. When studying solutions with null boundary in the BMS gauge, on the other hand, our results describe a new asymptotic
structure of this gravity theory. Similarly to the Maxwell algebra, the asymptotic generators
contain an additional set of Abelian generators $\mathcal{Z}_{m}$
which modifies the $\mathfrak{bms\mathrm{_{3}}}$ symmetry of Einstein gravity, to a deformed $\mathfrak{bms}\mathrm{_3}$ symmetry \cite{Caroca:2017onr},
where the additional central charge $c_{3}$ is proportional to the interaction constant $\alpha_{2}$.

It is worthwhile noticing that the extended solution in the BMS gauge that we found possesses several arbitrary functions of the boundary coordinates, which could be a manifestation of the existence of a number of asymptotic symmetries for this class of geometries. We have found the deformed $\mathfrak{bms}\mathrm{_3}$ algebra by setting all but one of them to zero. Therefore, an interesting question that arises in this analysis is that of the effect of other arbitrary functions in the asymptotic structure of the theory. If non-vanishing, would they lead to the same deformation of $\mathfrak{bms}\mathrm{_3}$ algebra or to a larger one?

There is also a relation with other algebras to be explored.
Remarkably, as shown in \cite{Caroca:2017onr}, the deformed $\mathfrak{bms}\mathrm{_{3}}$
algebra can alternatively be recovered as a flat limit of three copies
of the Virasoro algebra. On the other hand, the Maxwell symmetry can
be obtained as a flat limit of the AdS-Lorentz algebra. Taking into account these relations, a natural question is whether the three copies of the Virasoro symmetry describe an
asymptotic symmetry of a CS theory invariant under AdS-Lorentz algebra. This is a problem to be explored in near future. Similarly to
the Poincaré limit in the case of standard AdS Einstein gravity, a Maxwell limit can also be
applied to the AdS-Lorentz algebra. Therefore, it would be interesting
to analyze the flat Maxwell limit at the level of the asymptotic boundary
conditions. This is a work in progress. One could go even
further and generalize our results to the families of $\mathfrak{B\mathrm{_{k}}}$
and $\mathfrak{C\mathrm{_{k}}}$ CS gravities \cite{Concha:2016hbt}. In the context of holography, it would be natural to explore the duality between an asymptotically flat gravity theory and its holographic dual invariant under a deformation of $\mathfrak{bms}_3$ algebra, keeping in mind that the $\mathfrak{bms}_3$ algebra is isomorphic to the contracted conformal algebra, which is a symmetry of the holographic dual of the Rindler spacetime~\cite{Fareghbal:2014oba}.

Another aspect that deserves further investigation is the supersymmetric
extension of our results. One could expect that the asymptotic symmetry
of the $\mathcal{N}$-extended Maxwell CS supergravity corresponds
to a supersymmetric $\mathcal{N}$-extension of the deformed $\mathfrak{bms}\mathrm{_{3}}$, which is also a work in progress.

\newsection{Acknowledgment} The authors would like to thank R. Caroca,
L. Donnay, O. Fuentealba, J. Gomis, J. Matulich and R. Troncoso for valuable
discussions and comments.
This work was supported in parts by the Chilean
FONDECYT Projects N$^{\circ}$3170437, N$^{\circ}$1170765, N$^{\circ}$3170438,
N$^{\circ}$3160581 and N$^{\circ}$3160437, and the Grants VRIIP-UNAP N$^{\circ}$0114-17 and VRIEA-PUCV
N$^{\circ}$039.314/2018. NM was supported by a Becas-Chile postdoctoral grant of CONICYT.

\bibliographystyle{utphys}
\bibliography{ASM}

\end{document}